\documentstyle[12pt,openbib]{article}

\begin{document}



\title{Scattering of positronium from singly ionized helium and
appearing resonances.}

\author{P K Biswas \\
 Departamento de Fisica, Instituto
Tecnologico de Aeronautica, CTA\\
12228-900 S\~ao Jos\'e dos Campos, S\~ao Paulo, Brazil\\
and H. M. Gupta\\
Departamento de Fisica, IGCE, Universidade Estadual Paulista,\\
Rio Claro 13500-970, S\~ao Paulo, Brazil\\}

\maketitle

\begin{abstract}

The coupled-channel model [Phys. Rev. A {\bf 59}, 363 (1999)] that yields
converged low-energy phase shifts, and exact binding and resonance
features in Ps-H [Phys. Rev. A {\bf 59}, 2058 (1999)] is applied to study
Ps-He$^+$ scattering. Similar to PsH, resonances appear in the $S-, P-,
D-,$ etc partial waves in the Ps-He$^+$ system but in both the singlet
and triplet scattering channels. The latter signifies possible Rydberg
states of $e^+$ around singlet and triplet helium. The S-wave singlet
resonance at 2.79 eV agrees to that predicted earlier in $e^+$-He
entrance channel (energy difference 16.64 eV) at 19.27 eV [Chem. Phys.
Lett. {\bf 262}, 460 (1996)]. 





\end{abstract} 

\vskip 12cm

\newpage {\bf Introduction :} Ps-H is a fundamental few-body system
containing the positronium atom and has got an active theoretical
\cite{dr1,dr2,fr,ho1,ho2,ho3,w1,cpl,ab1,nimb,hg,sg,fs,bnm} and recent
experimental \cite{sdr} interest. In this article, we present an
investigation on ortho-Ps scattering from singly ionized helium atom. The
system is apparently similar to the four-body Ps-H system, but provides
the following additional interests. Firstly, within a similar four-body
framework, we can study Ps scattering dynamics relating to an ionic
target, where short-range exchange correlation is also present. So far,
Ps-ion scattering studies have been confined mostly to bare ions
\cite{ms,mt} (except a few studies \cite{ag,vh}) where there is no role
for the Pauli exclusion principle to play in absence of indistinguishable
fermions (electrons). Pauli exclusion principle introduces short-range
exchange forces which become crucial for Ps scattering. Because, in
absence of exchange, the potential in the elastic channel becomes exactly
zero due to charge and mass symmetry of Ps \cite{ft1}. Thus in order to
have a
comprehensive picture of the Ps-ion scattering dynamics, it is important
to investigate Ps scattering from an ionic target that has
indistinguishable fermions (electron or positron).

Secondly, we recall that, at low energies Coulomb interaction of
positronium atom (Ps) with hydrogen like system, leads to interesting
consequences like formation of resonances
\cite{dr1,dr2,fr,ho1,ho3,w1,cpl,ab1,nimb} and chemical binding
\cite{ho2,fs,bnm,sdr}. It has been interpreted in a Ps-H system that, the
positron forms Rydberg states with the residual H$^-$ ion \cite{dr2}. The
lowest order Rydberg state in the series lies below the Ps-H scattering
threshold and leads to chemical binding. Higher order states lie above
the scattering threshold and lead to resonances in the Ps-H continuum
\cite{ho3}. In Ps-H, the residual H$^-$ can exist only in the singlet
state and all these resonances and binding reported so far are in the
singlet state. For Ps-He$^+$ system, the positron shall find a residual
helium atom which can have both singlet and triplet spin configurations,
unlike the Ps-H case. So, in principle, in Ps-He$^+$, resonances could
appear in both the singlet and triplet scattering channels. Recently, a
metastable state for positronic helium ($e^+$He) is predicted \cite{rm}
with a binding energy of -2.2505916 Hatree and the state is found to be
stable under dissociation into PsHe$^+$ with a binding energy of
-0.0005916 Hatree. The chemical stability of PsHe$^+$ once again triggers
the speculation of resonances like those obtained in the PsH system.

As in Ps-H \cite{cpl,nimb}, we consider coupling of Ps-states to He$^+$
ion and investigate target elastic processes.  As He$^+$ is a tightly
bound system, we understand that target inelastic processes will not be
significant owing to its small polarizability compared to that of the Ps.
However, charge-transfer rearrangement like Ps+He$^+ \rightarrow e^++$He
is a relevant process and is expected to have its impact over
target-elastic and target-inelastic processes. However, we understand
that such channels are relevant in a complete {\it ab initio} calculation
where explicit electron-electron correlation and continuum effects are
essential to achieve convergence. Here, we are using regularized
exchange-correlation potentials which are able to simulate the
correlation and continuum effects \cite{cpl,ab1,nimb} (detailed in the
theory section). So, we abstain from considering such channels
explicitly. Since, their explicit inclusion in the present model may lead
to overcompleteness of the Hilbert space.

{\bf  Theory:} The theoretical formulation of
the Ps-He$^+$ scattering is very similar to that of the Ps-H system
except the difference comes from the nuclear charge, in the interaction
potential and target wave
function. With an antisymmetrized total wave function
\cite{w1,ab1,hg}, the momentum space Lippmann-Schwinger
equation, for a particular electronic spin state $S$, takes the
following form \cite{hg,sg}.
\begin{eqnarray} \label{2} f^{ S}_{\nu'\mu',\nu\mu} ( {\bf k_f,k_i})&=&
{\cal B}^S _{\nu'\mu',\nu\mu }({\bf k_f,k_i}) \nonumber \\
&-&\frac{1}{2\pi^2} \sum_{\nu{''}}\sum_{\mu''} \int {d\bf k{''}}\frac
{{\cal B}^S _ {\nu'\mu',\nu''\mu{''}} ({\bf k_f,k{''}}) f^S
_{\nu{''}\mu'',\nu\mu} ({\bf k{''},k_i})} {k_{\nu{''}\mu{''}}^2-k{''}^2+
i0} \end{eqnarray} 
where ${\bf
k}_i$, ${\bf k}_f$ are the initial and final momentum of the Ps atom with
respect to the center of mass and 
$k_{\nu{''}\mu{''}}^2=\frac{2m}{\hbar
^2}\{E-\epsilon_{\nu''}-\Upsilon_{\mu''}\}$; $m$, is the reduced mass of
Ps. $E$ represents the total energy of the system; $\epsilon_{\nu''}$ and
$\Upsilon_{\mu''}$ represent the binding energies of Ps and target
respectively. We need to solve two sets of coupled 
equations corresponding to singlet and triplet scattering
channels given by 
$S=0,1$. The
corresponding spin-averaged input potentials are given by
\begin{equation}\label{st}
{\cal B}^{0,1}_{\nu'\mu' \nu\mu}({\bf k_f,k_i}) = B^{D}_{\nu'\mu'
\nu\mu}({\bf k_f,k_i}) 
+ (-1)^{0,1} B^{E}_{\nu'\mu' \nu\mu}({\bf k_f,k_i})  
\end{equation} 
Where, $B^D$ and $B^E$ are the direct Born and Born-Oppenheimer
(BO) exchange 
amplitudes, respectively. The expression for $B^D$ can be
simplified and be
represented as \cite{ft2}:
\begin{eqnarray} \label{3} B^D_{\nu'\mu',\nu\mu} ({\bf k_f,
k_i})&= & -\frac{4}{q^2} <\phi_{\mu'}({\bf r_1})|2- e^{ i
{\bf q.r_1}}|\phi_\mu({\bf r_1})> \nonumber \\ &\times& <\chi_{\nu '}({\bf
t_{2}})|
e^{{i}{\bf q}.{\bf t_{2} }/2} - e^{-{i}{\bf q}.{\bf
t_{2} }/2}| \chi_{\nu}({\bf t_{2} })>, \end{eqnarray} 
where ${\bf q}= {\bf
k_i-k_f}$ and $\phi_\mu$'s and $\chi_\nu$'s are the eigenstates of
He$^+$ and Ps atom, respectively. ${\bf t}_j={\bf r}_j-{\bf
x}$; ${\bf x}$ and ${\bf r}_j$ are the coordinates of the positron
and the electron of the Ps atom.  
For $B^E$, we use the following non-local form \cite{nimb,ba1}, similar
to that applied  in PsH case:
\begin{eqnarray}\label{4}
B^{E}_{\nu'\mu',\nu\mu} ({\bf k_f,k_i})&= & \frac{4}{Q^2} <\phi_{\mu
'}({\bf r})| e^{{i} {\bf q.r}}|\phi_\mu({\bf r})> \nonumber \\ &\times&
<\chi_{\nu'}({\bf t})|e^{{i}{\bf q}.{\bf t}/2}|\chi_{\nu}({\bf t})>,
\end{eqnarray} where, $Q^2=(k_f^2 +k_i^2)/8+C^2[(\alpha^2_{\mu'}+
\alpha^2_\mu)/2+(\beta_ {\nu'}^2+\beta_\nu^2)/2]$.  
This form is arrived at by making use of the effective following type
of transformations to the exact BO expression for excahnge:
\begin{eqnarray}
\int \chi_{\nu'}({\bf x}-{\bf r}_2) \frac{1}{|{\bf r}_1-{\bf r}_2|}d^3r_2
&\approx& \frac{1}{(C\beta_{\nu'})^2} \chi_{\nu'}({\bf x}-{\bf r}_1)\\
\int \phi_{\mu}({\bf r}_2) \frac{1}{|{\bf r}_1-{\bf r}_2|}d^3r_2
&\approx& \frac{1}{(C\alpha_{\mu})^2} \phi_{\mu}({\bf r}_1)
\end{eqnarray}
where
$\alpha_\mu$,
$\beta_\nu$ etc are
parameters of the Slater orbitals of He$^+$ and Ps,
respectively; with
$\alpha^2_\mu$, $\alpha^2_{\mu'}$ representing the binding energies
of He$^+$ in the initial and final states in Rydberg units, and
$\beta_\nu^2$, $\beta_{\nu'}^2$ are the binding energies of
the initial and final Ps states in atomic units (au), respectively. A
dummy parameter $C$, is introduced in the expression of $Q^2$, which
can be used to
simulate the factor
$C^2[(\alpha^2_{\mu'}+ \alpha^2_\mu)/2+(\beta_ {\nu'}^2+\beta_\nu^2)/2]$
close to the total ionization energy of the system in $au$ or
facilitate any variation to this as has been effectively 
done for several electron impact cases \cite{mori}. 
It is of significant advantage that at
asymptotic energies, both the
value of the parameter $C$ and the regularized form above, do
loose their significance  and the exchange
potential ($B^E$) and hence
the
solution
of the coupled-equations coalesce, respectively, with those of the exact
CC formalism. 
In the present
work, however, we do not indulge to any fitting and rather paralyze the
sensitivity of the parameter by
re-introducing it in the following way:  $Q^2_j=(k_f^2
+k_i^2)/8+ 
(\beta_{\nu'}^2+\beta_\nu^2)/2
+C^2[(\alpha^2_{\mu'}+ \alpha^2_\mu)/2]  $ and by fixing $C^2=0.5$
so that the expression $(\beta_{\nu'}^2+\beta_\nu^2)/2
+0.5[(\alpha^2_{\mu'}+ \alpha^2_\mu)/2]$ yields the average binding
energies (between initial and final channel) of the electron of Ps and
He$^+$ in $au$, in
which the calculation is performed.

{\bf Numerical Procedures :} The three-dimensional LS equations, for a
particular electronic-spin state ($S$) are decomposed to coupled
one-dimensional partial wave equations, which are then solved by the
method of matrix inversion \cite{hg,mt}. For the present ionic target, we
register a slow convergence with respect to partial wave contributions,
compared to the neutral atomic target. We had to employ as much as thirty
partial waves at around 50-60eV incident energies, which is almost twice
to that employed for Ps-H case. Also, the convergence is found to be slow
with respect to the mesh points that discretizes the kernel of the
coupled-equations. We take maximum $48$ mesh points to discretize the
kernel compared to a maximum $32$ points taken for Ps-H. Present results
are numerically convergent upto fourth significant digit.

{\bf Results and Discussions :} We first discuss the results for the
partial and total cross sections. In figure 1, we plot the angle
integrated partial cross sections containing elastic and inelastic
transitions of Ps. Elastic (thin solid curve), Ps(1s$\rightarrow$2s)
(dotted curve) and Ps(1s$\rightarrow$2p) (short-dashed curve) transition
cross sections are obtained by solving coupled equations using a
3-Ps-state expansion. Results for Ps(1s$\rightarrow nlm; 3\le n\le 6$),
discrete excitations (long-dashed curve) and ionization (dot-dashed
curve) cross sections are obtained employing a first Born approximation
including exchange. Total (target elastic) cross sections which are
constructed from the results of Ps-elastic and all Ps-inelastic cross
sections, are also plotted in this figure (thick solid curve). At low
energies, the elastic and total cross sections have a sharp fall compared
to those of the Ps-H system. Similar sharp fall in the cross section is
also observed in the variational calculations for Ps-He$^+$ and Ps-$p$
scatterings \cite{vh} and in the coupled-channel calculation for Ps-$p$
scattering \cite{mt}.

At low energies, Van Reeth and Humberston (VRH) \cite{vh}, employing a
two-channel Kohn variational method, observed a Ramsauer minimum for the
elastic cross sections near 0.02 $\small{au}$ of Ps energy and a gentle
rise thereafter. We do not observe any such minimum in our elastic cross
sections but otherwise observe a gross agreement with VRH. To understand
the difference we recently perform a similar two-channel {\it ab initio}
CC calculation employing exact exchange and considering the
charge-transfer reaction Ps+He$^+ \rightarrow e^++$He. In our preliminary
results \cite{bgmg} we also do not record any such Ramsauer minimum.
Further insight seems warranted regarding the minimum.

At medium to high energies, we can expect the He$^+$ to behave more like
a bare proton and we find our results to agree with those of Ps-$p$
scattering cross sections. At 61.2 eV, the lowest energy considered by
Ratnavelu et al \cite{rms}, the elastic and Ps(1s$\rightarrow$2p) cross
sections are given by 0.5068 $\pi a_0^2$ and 8.426 $\pi a_0^2$ while the
present predictions for them are 0.5184 $\pi a_0^2$ and 8.198 $\pi a_0^2$
respectively. For Ps(1s$\rightarrow$2s) the agreement is not so close but
fair; at 61.2 eV, the present prediction is 1.712 $\pi a_0^2$ compared to
1.166 of Ratnavelu et al \cite{rms}. While elastic and Ps(2p) cross
sections agree so well, the difference in the Ps(2s) result is not
clearly understood. Among all the partial cross sections, the ionization
of Ps dominates the most and this, along with other Ps-inelastic
processes, causes a maximum in the total cross section beyond the Ps
excitation and ionization thresholds (5.1-6.8 eV). This is a
characteristic obtained in almost all Ps scattering calculations
\cite{cpl,ab1,nimb,ba1,bam} and also observed in the measured data
\cite{lr1,lr2,lr3}.

For Ps-He$^+$, the major interest is about possible resonances, similar
to Ps-H. This has become more relevant after Ryzhik and Mitroy \cite{rm}
have found the theoretical evidence of a metastable Ps-He$^+$ state with
a small binding energy of -0.0005916 Hatree. So, with a possible molecule
like feature, which has been predicted recently in PsH \cite{st,bd},
multiple resonances could prevail in the Ps-He$^+$ system corresponding
to various excited levels. As in the cases of Ps-H \cite{ab1} and Ps-Li
\cite{pb1}, we notice that there does not exist any resonant structure at
the static exchange level and the resonances are manifested with the
introduction of Ps-excited states in the coupling scheme. In figure 1, we
plot the singlet and triplet scattering cross sections for a) $S$-wave,
b) $P-$wave, and c) $D$-wave for the three-Ps-state model. As expected,
multiple resonances are found in both the singlet and triplet scattering
channels with each singlet state resonance is accompanied by a
corresponding triplet state resonance. For Ps-H \cite{dr2} and Ps-Li
\cite{pb1}, it was assumed that the resonances are appearing due to
Rydberg states of the positron around the singlet state H$^-$ and Li$^-$
ions, respectively and so resonances appeared there in the singlet
channel only. However, in Ps-He$^+$, the positron can have Rydberg states
around both the singlet and triplet spin configurations of the helium
atom and so resonances were expected in both the channels. The fact that
each singlet state resonance is accompanied by a corresponding triplet
state resonance and they appear in pairs is quite consistent with the
energy spectrum of helium where each electronic singlet-state
configuration has its triplet counterpart.

For S-wave, the singlet state resonances appear at 2.79 eV, 4.2 eV, and
4.85eV, while the triplet state resonances appear at 3.28 eV, 4.51 eV and
4.94 eV (see fig-2a). The widths are approximately given by 0.07 eV, 0.04
eV and 0.02 eV for the singlet state resonances and 0.065 eV, 0.02 eV,
and 0.0001 eV, for the triplet state resonances, respectively. We note
that, the triplet channel resonances appear beyond the corresponding
singlet channel resonances, in the energy scale. In the spin-singlet
channel, the short-range potential is attractive and it constructively
combines with the long-range polarization potential, while for the
triplet channel, the short-range potential is repulsive and it makes a
destructive cancellation with the attractive polarization potential
arising in the calculation through inclusion of Ps($1s \rightarrow 2p$)
transition in the coupling. We understand that this could be the reason
that the triplet channel resonances are placed beyond the singlet channel
ones. The energy difference between corresponding singlet-triplet
resonance pairs is found to diminish (0.49 eV, 0.31 eV, and 0.09 eV)
gradually for higher order resonances. This is also consistent with the
fact that the energy difference between singlet-triplet levels, diminish
with higher excited states of helium. Similar features are also obtained
for the $P-$ and $D-$wave resonances. In all the cases we find three pair
of resonances below the first excitation threshold of Ps (at 5.1 eV).

From figure 2b, we see that the $P-$wave resonances appear at 3.55 eV,
4.64 eV, and 4.99 eV for the singlet state and at 3.65 eV, 4.71 eV, and
5.0 eV for the triplet state. The energy difference between
singlet-triplet pairs are given by 0.10 eV, 0.07 eV, and 0.01 eV,
respectively and are also diminuing with higher order resonances. For $D-$
wave resonances, the energy difference between a singlet-triplet pair
diminishes further and the singlet and triplet resonances almost overlap
with each other (see fig-2c). Their positions are given by 4.19 eV, 4.907
eV, and 5.096 eV for the singlet state and 4.2 eV, 4.909 eV, and 5.096 eV
for the triplet state, respectively.

Now, we focus our attention to search for matching of these resonances
to physical levels of the four-body Ps-He$^+$ system. It could be
mentioned here that, in Ps-H, such correspondence has
been found to exist \cite{bd}. The lowest order $S-$wave and $P-$wave
singlet channel resonances are obtained at 2.79 eV and 3.55 eV. It is of
interest to find that the lowest excited singlet $s-$state and $p-$state
of He exist at 2.908 eV (-58.292-(54.4-6.8)), and 3.47 eV
(-57.73-(-54.4-6.8)). So, it is quite reasonable to believe that the above
two resonances are the manifestations of the rearrangement levels
$e^+$-He(1s,2$^1s$ and $e^+$-He($1s2^1p$)  of the Ps-He$^+$ system.

Further, it is quite relevant to note that the above lowest $S-$wave
singlet resonance at 2.79 eV corresponds quite well to the $S-$wave
singlet channel resonance predicted in the $e^+-$He entrance channel at
19.27 eV by Adhikari and Ghosh \cite{ag}. The correspondence becomes
apparent when one takes into account the entrance channel energy
difference of 16.64 eV between the two systems.

Manifestation of additional resonances seems to be quite relevant with
the following new findings. Recently, Ryzhik and Mitroy \cite{rm} have
concluded that PsHe$^+$ can exist in a metastable state with a small
binding energy of -0.0005916 Hatree (resembling similarity to Ps-H). And
secondly, Saito \cite{st} and Biswas and Darewych \cite{bd} have recently
found that the PsH system might have a molecule like feature, in addition
to its proposed atom like feature \cite{dr2} of a moving positron around
a H$^-$ ion. So, the metastable PsHe$^+$ can lead to various excited
levels before dissociation into Ps and He$^+$ or $e^+$ and He. These
levels may well be reflected as Feshbach resonances in the scattering
dynamics of Ps-He$^+$ and the additional resonances are thus worth
further investigation.

{\bf Conclusion:} In conclusion, we have studied positronium (Ps)  
scattering from singly ionized helium atom employing the model that
yielded precise resonance positions and binding energies for PsH system.
We find new resonances in both the singlet and triplet scattering
channels. In the light of Ps-H and Ps-Li resonances, these are supposed
to be due to the manifestation of positronic Rydberg states around the
singlet and triplet states of the residual helium atom. We find strong
correlation between few of the low lying resonances and some
rearrangement levels of the four-body Ps-He$^+$ system.  Although the
S-wave resonance predicted earlier in the $e^+$He entrance channel
\cite{ag} is yet to be verified experimentally, we consider that the
present work delivers a noteworthy information that a completely
different type of investigation also reveal the same.  In addition, the
model predicts new resonances which seem to be of real interest because
of the following two recently found physical features. Recently, Ryzhik
and Mitroy \cite{rm} has reported a metastable state for the Ps-He$^+$
with a small binding energy of -0.0005916 Hatree. On the other hand,
Saito \cite{st} and Biswas and Darewych \cite{bd} have found that the
four-body Ps-H system can have a molecule like manifestation, in addition
to its proposed atom-like structure \cite{dr2}. So, a possible molecule
like feature for the metastable PsHe$^+$ system would certainy account
for many of these additional resonances.

The work is  supported by the Funda\c c\~ao de Amparo \`a Pesquisa
do Estado de S\~ao Paulo of Brazil through project no. 99/09294-8.

\newpage


\newpage

{\bf Figure Captions:}

{\bf Fig. 1:} Angle integrated partial cross sections and target elastic
total cross sections (in units of $\pi a_0^2$). Elastic (thin-solid
curve), Ps($1s \rightarrow 2s$) (dotted curve), and Ps ($1s \rightarrow
2p$)  (short-dashed curve) excitation cross sections using 3-Ps-state
coupling. Ps(1s$\rightarrow$ $n=3,4,5,6$) discrete excitations
(long-dashed curve) and Ps ionization (dot-dashed curve) cross sections
using first Born approximation with regularized exchange. Total cross
section is represented by the thick-solid curve.

{\bf Fig. 2:} Variation of a) $S-$wave, b) $P-$wave, and c)
$D-$wave singlet and triplet cross sections (in units of $\pi
a_0^2$) for Ps-He$^+$ scattering for 3-Ps-state coupling.


\begin{thebibliography}{99}
\bibitem{ft1} From a partial wave expansion of the Ps form
factor in Eqn.3, it can be verified that it becomes zero for
elastic and even parity transitions.
\bibitem{ft2} Formal expressions for the Born amplitudes and their
simplification scheme leading to the present form can be found in
ref.\cite{w0,bg}. In the present case, $Z=2$, the atomic number of
the target.
\end{thebibliography}
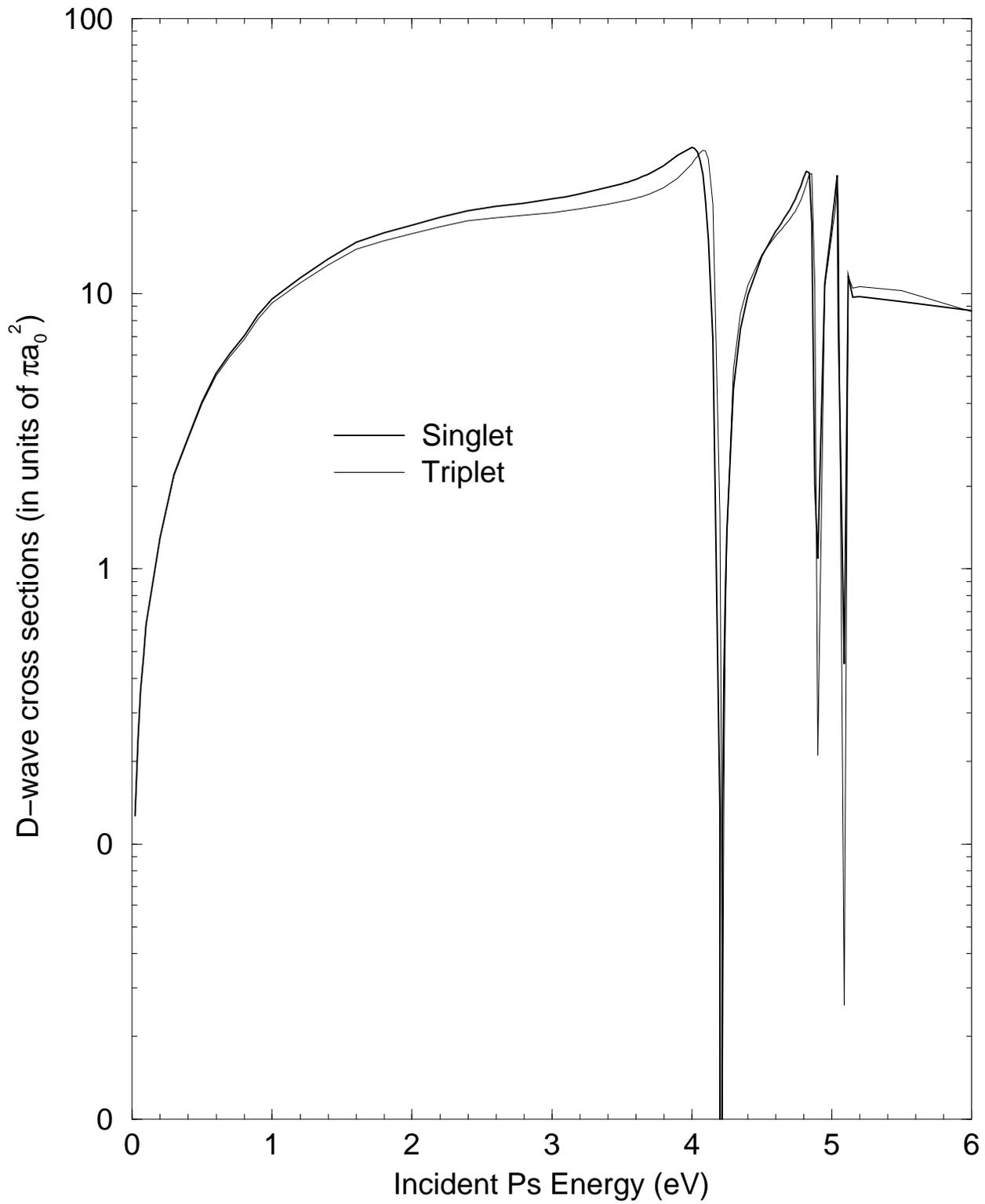
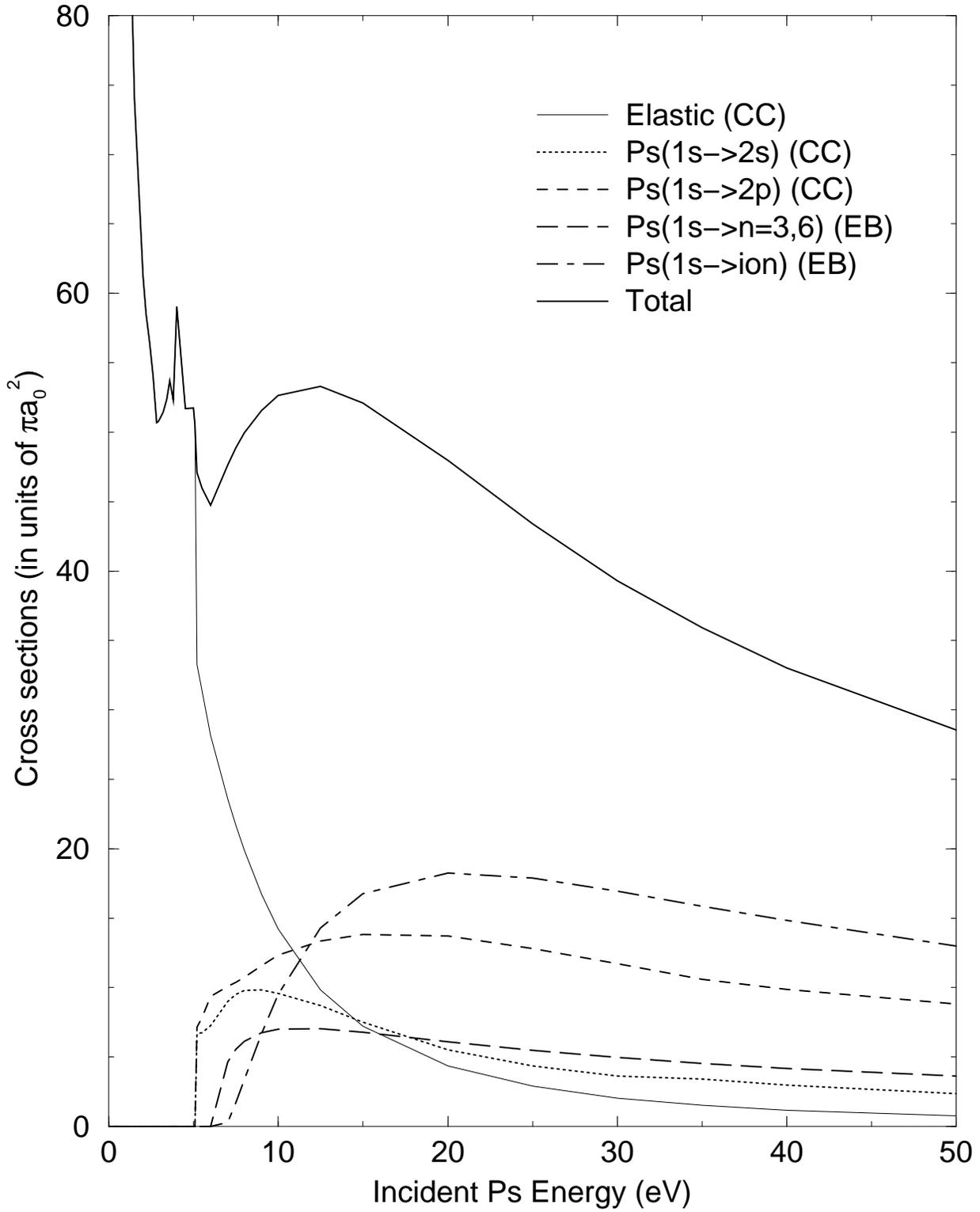
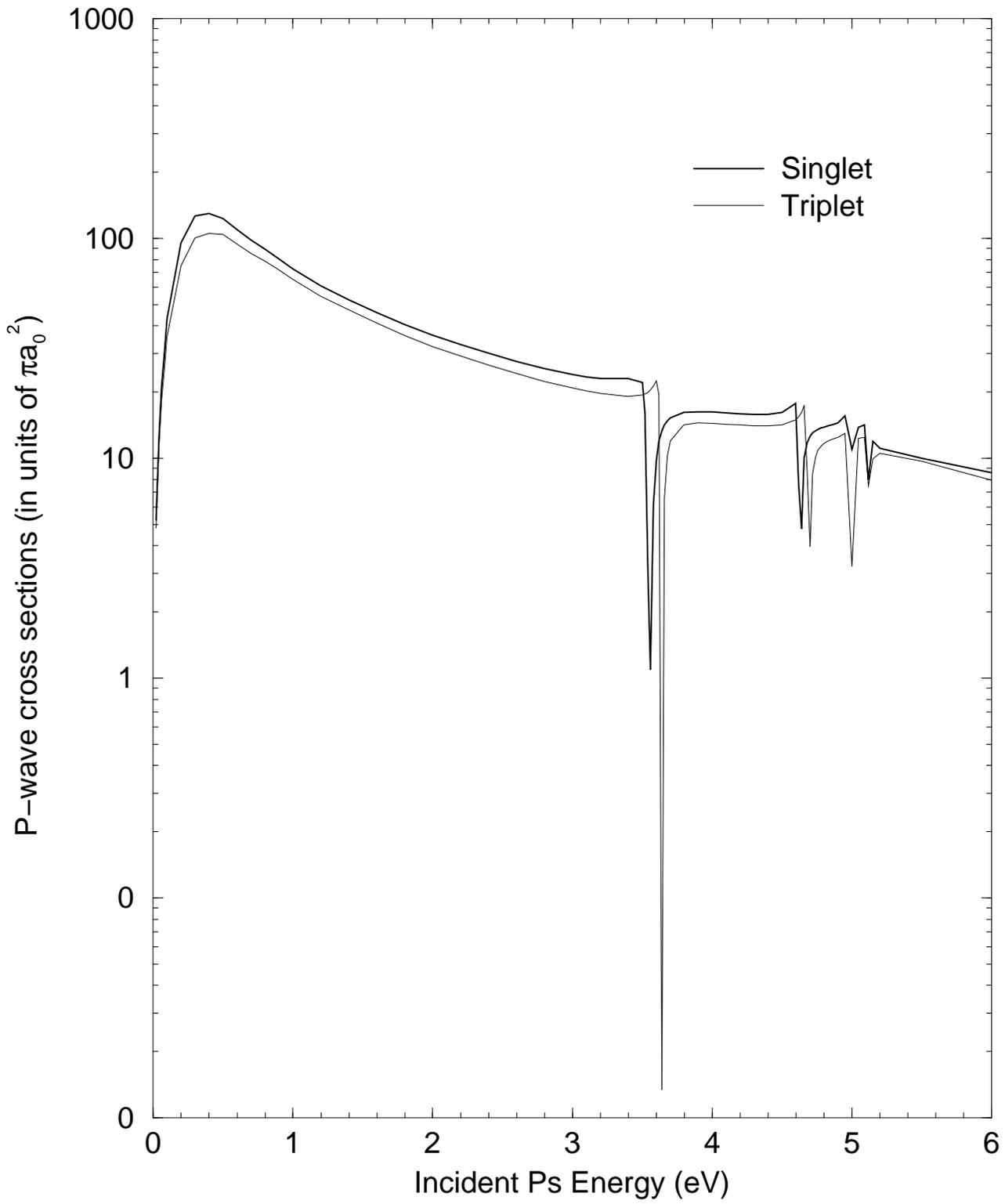
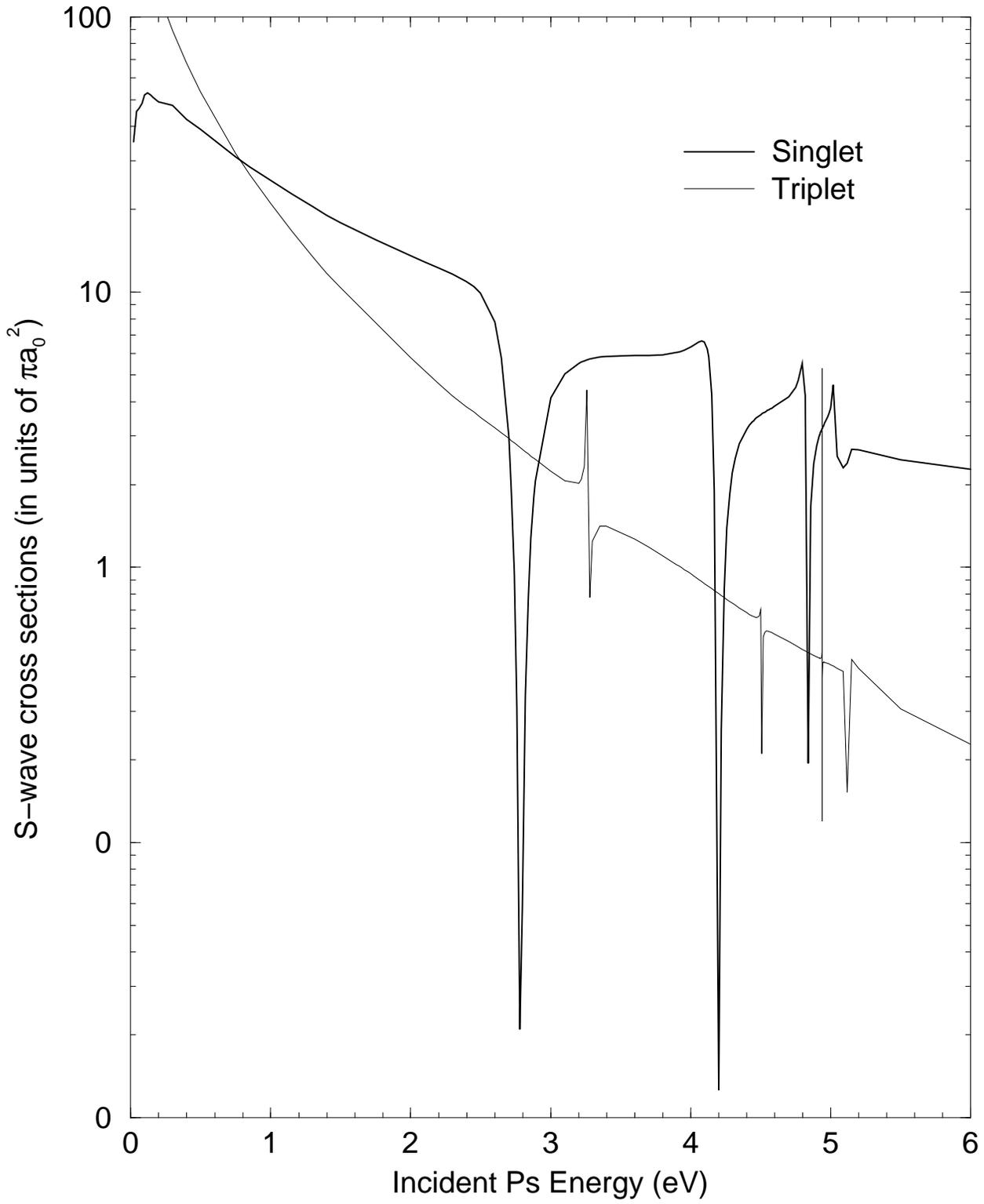
\end{document}